\documentclass[12pt]{iopart} 

\usepackage{epsfig} 
\usepackage{iopams}

\newcommand{\dt}{\Delta t} 
\newcommand{\tw}{t_\mathrm{w}} 

\newcommand{\eq}{\mathrm{eq}}
\newcommand{\C}{C}             
\newcommand{\X}{\chi}          
\newcommand{\Cc}{\mathcal{C}}  
\newcommand{\Xc}{\mathcal{X}}  

\newcommand{\In}{I}
\newcommand{\Hn}{H}
\newcommand{\Scal}{\mathcal{S}}
\newcommand{\Tcal}{\mathcal{T}}
\newcommand{\Hnhat}{\widehat{\Hn}}

\newcommand{\sgn}{\mathrm{sgn}}

\newcommand{\erfc}{\Phi}
\newcommand{\arccot}{\mathrm{arccot}}

\newcommand{\s}{\sigma}
\newcommand{\eps}{\varepsilon}

\begin{document} 
 
\title{Dynamic Heterogeneity in the Glauber-Ising chain}

\author{Peter Mayer$^{\dag,\ddag}$, Peter Sollich$^\ddag$, Ludovic
Berthier$^{\S,||}$ and Juan P. Garrahan$^{\P}$}

\address{$^\dag$Department of Chemistry, Columbia University, 
3000 Broadway, New York, NY 10027, US}

\address{$^\ddag$Department of Mathematics, King's College, Strand,
London, WC2R 2LS, UK}

\address{$\S$Theoretical Physics, University of Oxford, 1 Keble
Road, Oxford, OX1 3NP, UK}

\address{$||$Laboratoire des Collo\"{\i}des, Verres et Nanomat\'eriaux,
Universit\'e Montpellier II and UMR 5587 CNRS, 34095 Montpellier Cedex 5,
France} 

\address{$^\P$School of Physics and Astronomy, University of
Nottingham, Nottingham, NG7 2RD, UK}

\begin{abstract} 
In a recent paper [P. Mayer et al., Phys. Rev. Lett.  {\bf 93}, 115701 
(2004)] it was shown, by means of experiments, theory and
simulations, that coarsening systems display dynamic heterogeneity
analogous to that of glass formers.  Here, we present a detailed
analysis of dynamic
heterogeneities in the Glauber-Ising chain.  We discuss how dynamic
heterogeneity in Ising systems must be measured through connected
multi-point correlation functions.  We show that in the coarsening
regime of the Ising chain these multi-point functions reveal the
growth of spatial correlations in the dynamics, beyond what can be
inferred from standard two-point correlations. They have non-trivial
scaling properties, which we interpret in terms of the
diffusion-annihilation dynamics of domain walls.
In the equilibrium dynamics of the
Ising chain, on the other hand, connected multi-point functions vanish
exactly and dynamic heterogeneity is not observed.  Our results
highlight the similarities between coarsening systems and glass
formers.
\end{abstract} 

\maketitle

\section*{Introduction}
\label{sec:intro}

An obvious question to ask about glass-forming liquids is whether the
dramatic slow-down of the dynamics on cooling is correlated with a 
corresponding increase in an appropriately defined length scale. 
Critical slowing-down around second-order phase
transitions, for example, is correlated with the divergence of a
static correlation length.  In supercooled liquids and glasses, the
consensus is that there is no growing {\em static} length scale, since
the static structure---as measured, e.g., by the amplitude of
density fluctuations---changes only negligibly while relaxation
time scales grow by orders of magnitude.  Any growing length scale in
glassy systems must therefore reflect the spatial structure of the
{\em dynamics}.  In order for this spatial structure to be
non-trivial, the dynamics must vary from point to point: it must be
{\em heterogeneous}.  The simplest conceptual picture of such
dynamical heterogeneity is that some regions of a material are fast
and others slow.  The identities and locations of these regions may of
course change over time.

We will not try to review the literature on dynamical
heterogeneity, which is vast, and refer instead
to~\cite{Ediger00,Sillescu99,GlotzerReview}.  We focus in this paper
on the characterization of dynamical heterogeneities via multi-point
correlations~\cite{Diezemann03,DonFraGloPar02,FraDonParGlo99,FraPar00, 
GC02,Berthier03b,BerGar03,BiroliBouchaud,JacBerGar05}.
In the context of lattice models, one considers a spatial correlation
function between sites $i$ and $j$, of the general form
$\C_{ij}=\langle F_i F_j\rangle - \langle F_i \rangle \langle
F_j\rangle$, but with $F_i$ {\em itself} a two-time quantity such as
$F_i = A_i(t)A_i(\tw)$ and $A$ some local observable. Thus $\C_{ij}$
is a four-point correlation function. It essentially measures how the 
relaxation of $A$ at point $i$ is correlated with the relaxation at 
point $j$. Such a definition should therefore pick up dynamical 
heterogeneities. Associated with $C_{i,j}$ is a so-called susceptibility, 
i.e., the spatially integrated correlation $\X = (1/N)\sum_{ij} \C_{ij}$: 
if the length scale on which the dynamics is correlated 
grows, then so should this fourth-order susceptibility.

It is worth noting that the four-point susceptibility
$\X=(1/N)\sum_{ij} \C_{ij}$ can be re-expressed as $\X = (1/N) 
\left[\langle q^2 \rangle - \langle q \rangle^2 \right]$ where 
$q=\sum_i A_i(t)A_i(\tw)$ measures the ``overlap'' between 
configurations at time $\tw$ and $t$. From this we can easily gain 
qualitative insight into the dependence of $\X$ on $\dt = t - \tw$. 
At $\dt=0$, one does not
expect large fluctuations in $q$; indeed, in a spin system and for
$A_i=\s_i$, $q$ is a constant for $t=\tw$ and hence $\X=0$. If, in the
opposite limit $\dt\to\infty$, the system decorrelates from its state
at time $\tw$, then one expects $q$ to decay to a small value, again
with only minor fluctuations. For intermediate times, however, the
fact that a glass
may remain trapped near the $\tw$-configuration for a time that is
both long and strongly dependent on the state at $\tw$ leads to
large fluctuations of $q$ between different dynamical histories and
therefore to a large value of $\X$.

The usefulness of four-point correlations for understanding heterogeneous 
dynamics motivates us to consider their behaviour in {\em coarsening} 
systems~\cite{Diezemann03}. Because -- contrary to glass formers 
-- coarsening systems develop strong spatial correlations 
at long times, appropriately modified four-point correlations 
must be considered. It was shown in a recent letter \cite{PeterLetter} 
that these reveal non-trivial spatio-temporal correlations in
coarsening systems which show strong analogies with aging glasses.
This is, in fact, also the case for the Glauber-Ising chain, 
which we study in this paper. Its analytical tractability 
makes it an interesting candidate for exploring four-point 
correlations in coarsening dynamics. Results were partially 
announced in \cite{PeterLetter}. 

Explicitly, the standard four-point function that is studied in the 
literature on dynamical heterogeneities becomes for our spin system 
\begin{equation}
  \C_{l-k}(\dt,\tw) = 
    \langle \s_k(t) \s_k(\tw) \s_l(t) \s_l(\tw) \rangle - 
    \langle \s_k(t) \s_k(\tw) \rangle \langle \s_l(t) \s_l(\tw) \rangle,
  \label{equ:Cnregdef}
\end{equation}
where $t=\dt+\tw\ge\tw\ge 0$ and the times $t,\tw$ are measured from
the quench. The definition (\ref{equ:Cnregdef}) implies that
$\C_n(\dt,\tw)$ is an even function of $n$. At $\dt=0$ it in
fact vanishes for all $n$ since Ising spins satisfy $\sigma^2 = 1$. 
In the opposite limit $\dt\to\infty$, because the configurations
at $\tw$ and $t$ will decorrelate, $\C_{l-k}(\dt,\tw)$ approaches $\langle
\s_k(\tw) \s_l(\tw) \rangle \langle \s_k(t) \s_l(t) \rangle - \langle
\s_k(t) \rangle \langle \s_k(\tw) \rangle \langle \s_l(t) \rangle
\langle \s_l(\tw) \rangle$. The last term vanishes, so this limit
reduces to the product of the spatial correlations at times $\tw$ 
and $t\to\infty$. This argument holds quite generally in a spin system
without an overall magnetization.  However, in a typical glassy system
the spatial correlations at times $\tw$ and $t$ will be
comparable and of limited range.  In a coarsening system, on the other
hand, the spatial correlations at time $t$ have a diverging range as
$t\to\infty$, with $\langle \s_k(t) \s_l(t) \rangle\to 1$, unless 
specific symmetries of the Hamiltonian are present~\cite{JacBerGar05}. 
This suggests that the large $\dt$ limit of the standard four-point 
correlation $\C_n(\dt,\tw)$ will be larger in a
coarsening system than in glasses, with the growth of spatial 
two-point correlations obscuring genuine four-point correlation
effects.

This is indeed what we will find: to see genuine four-point
correlations, we need to consider the {\em ``connected''} four-point
correlation
\begin{eqnarray}
  \fl \Cc_{l-k}(\dt,\tw) = 
  \langle \s_k(t) \s_k(\tw) \s_l(t) \s_l(\tw) \rangle 
  -\langle \s_k(t) \s_k(\tw) \rangle \langle \s_l(t) \s_l(\tw) \rangle 
  \nonumber \\
  -\langle \s_k(t) \s_l(t) \rangle \langle \s_k(\tw) \s_l(\tw) \rangle 
  +\langle \s_k(t) \s_l(\tw) \rangle \langle \s_k(\tw) \s_l(t) \rangle, 
  \label{equ:Cndef}
\end{eqnarray}
which differs from the standard version by the terms in the second
line. One observes that the first of these just cancels the residual
term from the four-point average in the limit $\dt\to \infty$. We
therefore expect that $\Cc_n(\dt,\tw)\to 0$ for large $\dt$: the
connected definition eliminates the uninteresting contributions from
spatial two-point correlations. The second term in the second line of
(\ref{equ:Cndef}) can then be motivated as compensating for the first
one at short times $\dt$, ensuring that, like $\C_n$, $\Cc_n$ vanishes at
$\dt=0$. We note finally that $\Cc_n$ is even in $n$ as was the case
for $\C_n$; at $n=0$, the definition (\ref{equ:Cndef}) in fact
implies that $\Cc_0=0$ at all times.

We will also consider the four-point susceptibilities
associated with (\ref{equ:Cnregdef}), (\ref{equ:Cndef}). 
These are defined as 
\begin{equation}
  \X(\dt,\tw)=\sum\limits_{n=-\infty}^{\infty} \C_n(\dt,\tw) 
  \quad \mbox{and} \quad 
  \Xc(\dt,\tw)=\sum\limits_{n=-\infty}^{\infty} \Cc_n(\dt,\tw).
  \label{equ:Cdef}
\end{equation}

The layout of this paper is as follows. First we derive an exact 
expression for the connected four-point correlation $\Cc_n(\dt,\tw)$ in 
Section~\ref{sec:derivation}. In Section~\ref{sec:1deq} the equilibrium 
behaviour of the standard and connected four-point functions is discussed. 
Scalings of the connected four-point correlation and its associated 
susceptibility for non-equilibrium coarsening dynamics are analysed in 
Section~\ref{sec:1dneq}. We then interpret our results in 
Section~\ref{sec:randomwalk} in terms of the random-walk dynamics of 
domain walls. The behaviour of the standard four-point functions is briefly 
presented in Section~\ref{sec:regnoneq}. We conclude in the final Section.

\section{Derivation of the Connected Four-Point Correlation}
\label{sec:derivation}

In this section we analyse the dynamics of the 
Glauber-Ising chain~\cite{Glauber63}, quenched from a random 
initial configuration to some temperature $T \geq 0$. To recap 
briefly, the model has Hamiltonian $\mathcal{H} = -\sum_i 
\sigma_i \, \sigma_{i+1}$, where $\sigma_i=\pm1$ ($i=1,\ldots N$) 
are $N$ Ising spins subject to periodic boundary conditions. 
Glauber dynamics consists in each spin $\sigma_i$ flipping with rate 
$\frac{1}{2}[1-\frac{1}{2}\gamma\sigma_i(\sigma_{i-1}+\sigma_{i+1})]$, 
where $\gamma = \tanh(2/T)$. 

General expressions for two-time multispin correlation functions 
in the Glauber-Ising chain are given in~\cite{MaySol04}, for the 
finite model quenched at $t=0$ from equilibrium at an initial temperature 
$T_\mathrm{i} > 0$ to arbitrary $T \geq 0$. Let us now recall 
some results relevant for the present analysis: in the thermodynamic 
limit $N\to\infty$ and for a quench from a random initial state 
$T_\mathrm{i} \to \infty$ we have the following representations
\begin{eqnarray}
  \fl \langle \s_i(t) \s_j(\tw) \rangle = 
    \rme^{-\dt} \In_{i-j}(\gamma \dt) + \mathcal{E}_{i,j}^{\,(j)}, 
    \label{equ:cor11} \\
  \fl \langle \s_{i_1}(t) \s_{i_2}(t) \s_{j_1}(\tw) \s_{j_2}(\tw) \rangle =
    \big[\mathcal{F}_{i_1,i_2}^{\,(j_1,j_2)} + \Hn_{i_2-i_1}(2\dt)\big] \, 
    \Hn_{j_2-j_1}(2\tw) 
    \nonumber \\
  -\big[+\rme^{-\dt} \In_{i_1-j_1}(\gamma \dt) + 
    \mathcal{E}_{i_1,j_1}^{\,(j_1,j_2)} \big] 
    \big[-\rme^{-\dt} \In_{i_2-j_2}(\gamma \dt) + 
    \mathcal{E}_{i_2,j_2}^{\,(j_1,j_2)} \big] 
    \nonumber \\
  +\big[-\rme^{-\dt} \In_{i_1-j_2}(\gamma \dt) + 
    \mathcal{E}_{i_1,j_2}^{\,(j_1,j_2)} \big] 
    \big[+\rme^{-\dt} \In_{i_2-j_1}(\gamma \dt) + 
    \mathcal{E}_{i_2,j_1}^{\,(j_1,j_2)} \big], 
    \label{equ:cor22} 
\end{eqnarray}
for two and four-spin two-time correlation functions. In (\ref{equ:cor22}) 
the indices must satisfy $i_1<i_2$ and $j_1 < j_2$. The general form of 
the functions $\mathcal{E}$ and $\mathcal{F}$ is 
\begin{eqnarray}
  \fl \mathcal{E}_{i_\eps,j_\nu}^{\,\bi{j}} = 
  \sum_p \left[\prod_{\lambda=1}^{\mathrm{dim}(\bi{j})} \sgn(j_\lambda-p) 
  \right] 
  \rme^{-\dt} \, \In_{i_\eps-p}(\gamma \dt) \,
  \Hn_{j_\nu-p}(2\tw), 
  \label{equ:Esum} \\
  \fl \mathcal{F}_{i_\eps,i_\delta}^{\,\bi{j}} = 
  \sum_{p,q} \left[ \prod_{\lambda=1}^{\mathrm{dim}(\bi{j})} \sgn(j_\lambda-p) 
  \, \sgn(j_\lambda-q) \right] 
  \rme^{-2 \dt} \, \In_{i_\eps-p}(\gamma \dt) \, \In_{i_\delta-q}(\gamma \dt) 
  \,\Hn_{q-p}(2\tw). 
  \label{equ:Fsum}
\end{eqnarray}
In (\ref{equ:Esum}) and (\ref{equ:Fsum}) the products over the sign-functions 
$\sgn(x)$, satisfying $\sgn(0)=0$ and $\sgn(x) = \frac{x}{|x|}$ otherwise, 
involve the indices of {\em all} spins at the earlier time $\tw$; so 
when substituting (\ref{equ:Esum}) into (\ref{equ:cor11}) there is only 
one factor $\sgn(j-p)$ while in (\ref{equ:cor22}) we have two factors 
$\sgn(j_1-p) \, \sgn(j_2-p)$. The summations over $p,q$ in 
(\ref{equ:Esum}), (\ref{equ:Fsum}) are to be taken over the entire 
chain $-\infty < p,q < \infty$. Finally, the functions $\In_q(t)$ 
denote modified Bessel functions~\cite{Mathbook} while the $\Hn_q(t)$ 
have the representation~\cite{MaySol04}
\begin{equation}
  \Hn_q(t) = \frac{\gamma}{2} 
  \int_0^t \rmd\tau \, \rme^{-\tau} 
  \left[\In_{q-1}(\gamma\tau) - \In_{q+1}(\gamma\tau) \right]. 
  \label{equ:HI}
\end{equation}
The physical meaning of $\Hn$ is 
\begin{equation}
  \langle \s_i(t) \s_j(t) \rangle = \Hn_{j-i}(2t), 
  \label{equ:structfact}
\end{equation}
but this holds for $i<j$ only: in contrast to the two-spin correlation 
(\ref{equ:structfact}), the function $\Hn_q(t)$ is odd in $q$ 
and zero for $q=0$. Further properties of $H$, which are summarized 
in~\cite{MaySol04}, are recalled below as and when required.

Let us now focus on the connected two-time correlation $\Cc_{l-k}(\dt,\tw)$ 
defined in (\ref{equ:Cndef}). In order to be able to express the 
four-spin term using (\ref{equ:cor22}) we require 
$k<l$; below, $n$ always stands for $l-k$ and is assumed 
to be positive. Also substituting 
(\ref{equ:cor11}) and (\ref{equ:structfact}) for the corresponding 
two-spin correlations gives, after some rearranging, 
\begin{eqnarray}
  \fl \Cc_n(\dt,\tw) = 
    \big[ \mathcal{F}_{k,l}^{\,(k,l)} + \Hn_n(2\dt) - \Hn_n(2t) \big] 
    \Hn_n(2\tw) \nonumber \\
  - \big[ \mathcal{E}_{k,k}^{\,(k)} - \mathcal{E}_{k,k}^{\,(k,l)} \big] 
    \big[ \mathcal{E}_{k,k}^{\,(k)} + \mathcal{E}_{k,k}^{\,(k,l)} + 
    2 \,\rme^{-\dt} \,\In_0(\gamma \dt) \big] \nonumber \\
  + \big[ \mathcal{E}_{k,l}^{\,(l)}\, + \mathcal{E}_{k,l}^{\,(k,l)} \big] 
    \big[ \mathcal{E}_{k,l}^{\,(l)}\, - \mathcal{E}_{k,l}^{\,(k,l)} + 
    2 \,\rme^{-\dt} \,\In_n(\gamma \dt) \big]. 
   \label{equ:CnEF}
\end{eqnarray}
Here we have used 
$\mathcal{E}_{l,l}^{\,(l)} = \mathcal{E}_{k,k}^{\,(k)}$, 
$\,\mathcal{E}_{l,k}^{\,(k)} = \mathcal{E}_{k,l}^{\,(l)}$, 
$\,\mathcal{E}_{l,l}^{\,(k,l)} = -\mathcal{E}_{k,k}^{\,(k,l)}$ and 
$\mathcal{E}_{l,k}^{\,(k,l)} = -\mathcal{E}_{k,l}^{\,(k,l)}$. These 
properties follow directly from the definition (\ref{equ:Esum}) of 
$\mathcal{E}$ and reflect symmetries like $\langle \sigma_k(t) 
\sigma_l(\tw) \rangle = \langle \sigma_l(t) \sigma_k(\tw) \rangle$. 
The problem of analysing $\Cc_n(\dt,\tw)$ is now reduced to rewriting 
(\ref{equ:CnEF}) in a convenient form. To this end one could utilize 
the closed representations for $\mathcal{E}$ and $\mathcal{F}$ derived 
in \cite{MaySol04}. These were, however, constructed for cases where 
the spins at the earlier time $\tw$ are at a fixed and small distance. 
In the current context this distance is given by $n$ and we are interested 
in studying the scaling behaviour for $n\to\infty$ or working out the 
infinite sum over $n$ in (\ref{equ:Cdef}). It is therefore necessary to 
develop a new approach for dealing with the expression (\ref{equ:CnEF}). 

As we show in the following it is convenient to rearrange the sums 
$\mathcal{E}$ and $\mathcal{F}$. Let us first 
consider the sums $\mathcal{E}$, appearing in (\ref{equ:CnEF}) only 
in very particular combinations. After substitution of (\ref{equ:Esum}) 
and a shift in the summation variable we obtain, for instance, 
\begin{equation}
  \mathcal{E}_{k,k}^{\,(k)} \pm \mathcal{E}_{k,k}^{\,(k,l)} = 
  \sum_p \left[ 1 \pm \sgn(n+p) \right] \sgn(p) \, \rme^{-\dt} \, 
  \In_p(\gamma \dt) \, \Hn_p(2\tw).
  \label{equ:Epm}
\end{equation}
This sum obviously reduces to a semi-infinite one due to the factor 
in the square brackets, for either choice of the sign. 
In order to lighten the notation and for the subsequent analysis it 
is convenient to introduce the weight function 
\begin{equation}
  w_q = \left\{ 
    \begin{array}{rcl} 
      1 &                        & q=0 \\ 
      2 & \quad \mbox{for} \quad & 1 \leq q < n \\ 
      1 &                        & q=n 
    \end{array} \right. \ . 
  \label{equ:wq}
\end{equation}
We also use a slight modification of $\Hn$ 
\begin{equation}
  \Hnhat_q(t) = \delta_{q,0} + \Hn_q(t), 
  \label{equ:Hnhat}
\end{equation}
and define 
\begin{eqnarray}
  \Scal_q  = \rme^{-\dt} \, \In_0(\gamma \dt) \, \Hnhat_q(2\tw) + 
    2 \sum_{p=1}^{\infty} \rme^{-\dt} \, \In_p(\gamma \dt) \, \Hn_{q+p}(2\tw), 
    \label{equ:Ssum} \\
  \Tcal_q  = \rme^{-\dt} \, \In_n(\gamma \dt) \, \Hnhat_q(2\tw) + 
    2\sum_{p=1}^{\infty} \rme^{-\dt} \, \In_{n+p}(\gamma \dt)\,\Hn_{q+p}(2\tw).
    \label{equ:Tsum} 
\end{eqnarray}
In terms of (\ref{equ:wq}-\ref{equ:Tsum}) the two combinations of 
$\mathcal{E}$'s in (\ref{equ:Epm}) can then easily be shown to equal 
\begin{eqnarray}
  \mathcal{E}_{k,k}^{\,(k)} - \mathcal{E}_{k,k}^{\,(k,l)} = 
    \Tcal_n, \label{equ:EminuskkST} \\
  \mathcal{E}_{k,k}^{\,(k)} + \mathcal{E}_{k,k}^{\,(k,l)} + 2\, \rme^{-\dt} \, 
  \In_0(\gamma \dt) = 
    \Scal_0 + \sum_{q=0}^n w_q \, \rme^{-\dt} \, \In_q(\gamma \dt) \, 
    \Hnhat_q(2\tw). \label{equ:EpluskkST}
\end{eqnarray}
Similarly we find for the other two combinations of $\mathcal{E}$'s 
in (\ref{equ:CnEF}) 
\begin{eqnarray}
  \mathcal{E}_{k,l}^{\,(l)} + \mathcal{E}_{k,l}^{\,(k,l)} = 
    \Scal_n, \label{equ:EplusklST} \\
  \mathcal{E}_{k,l}^{\,(l)} - \mathcal{E}_{k,l}^{\,(k,l)} + 2\, \rme^{-\dt} \, 
  \In_n(\gamma \dt) = 
    \Tcal_0 + \sum_{q=0}^n w_q \, \rme^{-\dt} \, \In_{n-q}(\gamma \dt) \, 
    \Hnhat_q(2\tw). \label{equ:EminusklST}
\end{eqnarray}
Next consider the double sum $\mathcal{F}$. Here the relevant combination 
is $\mathcal{F}_{k,l}^{\,(k,l)} + \Hn_n(2\dt) - \Hn_n(2t)$. Based on the 
identity $\In_q(x+y) = \sum_p \In_p(x) \, \In_{q+p}(y)$ and (\ref{equ:HI}) 
one verifies that 
\begin{equation}
  \Hn_{l-k}(2t) - \Hn_{l-k}(2\dt) = 
  \sum_{p,q} \rme^{-2\dt} \, 
  \In_{k-p}(\gamma \dt) \, \In_{l-q}(\gamma \dt) \, \Hn_{q-p}(2\tw).
  \label{equ:Hnaux}
\end{equation}
Expressing $\mathcal{F}$ via (\ref{equ:Fsum}) and using (\ref{equ:Hnaux}) 
then yields
\begin{eqnarray}
  \fl  \mathcal{F}_{k,l}^{\,(k,l)} + \Hn_n(2\dt) - \Hn_n(2t) = 
    - \sum_{p,q} \left[1 - \sgn(k-p)\,\sgn(k-q)\,\sgn(l-p)\,\sgn(l-q) \right]
    \nonumber \\
  \times \rme^{-2\dt} \, \In_{k-p}(\gamma \dt) \, 
    \In_{l-q}(\gamma \dt) \, \Hn_{q-p}(2\tw). 
    \label{equ:Fdiff} 
\end{eqnarray}
In analogy with (\ref{equ:Epm}) the factor in the square bracket in 
(\ref{equ:Fdiff}) is non-zero only in a restricted range of the summation 
variables $p,q$. 
In fact, the two-dimensional 
infinite sum in (\ref{equ:Fdiff}) may be rewritten as a finite number 
of one-dimensional semi-infinite sums. From this procedure, which 
is slightly cumbersome but trivial, and using the 
notation (\ref{equ:wq} - \ref{equ:Tsum}) we obtain
\begin{equation}
  \fl \mathcal{F}_{k,l}^{\,(k,l)} + \Hn_n(2\dt) - \Hn_n(2t) = 
    \sum_{q=0}^n w_q \, \rme^{-\dt} \, \In_q(\gamma \dt) \, \Tcal_q - 
    \sum_{q=0}^n w_q \, \rme^{-\dt} \, \In_{n-q}(\gamma \dt) \, \Scal_q.
    \label{equ:FST}
\end{equation}
In terms of equations (\ref{equ:EminuskkST}-\ref{equ:EminusklST}) and 
(\ref{equ:FST}) our representation (\ref{equ:CnEF}) for the connected 
four-point correlation $\Cc_n(\dt,\tw)$ is thus transformed into 
\begin{eqnarray}
  \fl \Cc_n(\dt,\tw) = 
    \Scal_n \, \Tcal_0  - \Tcal_n \, \Scal_0 
    \nonumber  \\
  + \sum_{q=0}^n w_q \, \rme^{-\dt} \, \In_q(\gamma \dt) 
    \big[ \Tcal_q \, \Hn_n(2\tw) - 
    \Hnhat_q(2\tw) \, \Tcal_n \big] 
    \nonumber \\
  - \sum_{q=0}^n w_q \, \rme^{-\dt} \, \In_{n-q}(\gamma \dt) 
    \big[ \Scal_q \, \Hn_n(2\tw) - 
    \Hnhat_q(2\tw) \, \Scal_n \big]. 
    \label{equ:CnST} 
\end{eqnarray}
Equation~(\ref{equ:CnST}) forms the basis for our subsequent analysis 
of $\Cc_n(\dt,\tw)$. Note that up to this point we have not carried out 
any summations, except for (\ref{equ:Hnaux}). The derivation of 
(\ref{equ:CnST}) relies only on direct cancellations occurring 
in (\ref{equ:CnEF}).

\section{Equilibrium}
\label{sec:1deq}

A striking feature of (\ref{equ:CnST}) is that it allows us to study 
the equilibrium behaviour of $\Cc_n(\dt,\tw)$, still without working out 
any sums at all. To see this we first notice that in equilibrium we have, 
as can be shown~\cite{MaySol04} from (\ref{equ:HI}) and (\ref{equ:Hnhat}), 
\begin{equation}
  \Hnhat_q^{\eq} = \lim_{\tw \to \infty} \Hnhat_q(2\tw) =  
  \xi^q \quad \mbox{for} \quad q\geq0, 
  \label{equ:Heq}
\end{equation}
where 
\begin{equation}
  \xi = \frac{1-\sqrt{1-\gamma^2}}{\gamma}=\tanh(1/T). 
  \label{equ:xi} 
\end{equation}
Due to the exponential dependence of $\Hnhat_q^{\eq}$ on $q$, the expressions 
(\ref{equ:Ssum}), (\ref{equ:Tsum}) for $\Scal,\Tcal$ satisfy in equilibrium
\begin{eqnarray}
  \Scal_q^{\eq} = \xi^q \, \Scal_0^{\eq} \label{equ:Seq} 
  \quad \mbox{and} \quad 
  \Tcal_q^{\eq} = \xi^q \, \Tcal_0^{\eq} \label{equ:Teq}. 
\end{eqnarray}
It therefore follows immediately from (\ref{equ:CnST}) that our 
connected four-point correlation vanishes in equilibrium, i.e., 
\begin{equation}
  \Cc_n^{\eq}(\dt)=\lim_{\tw \to \infty} \Cc_n(\dt,\tw) = 0, 
  \label{equ:Cneq}
\end{equation}
for all $n$, $\dt \geq 0$ and at any temperature $T>0$ as specified
via $\gamma$. Looking back at the definition of $\Cc_n(\dt,\tw)$ in equation 
(\ref{equ:Cndef}), this implies an exact decomposition of four-spin 
{\em two-time} correlations into pairwise correlations. In other words, 
in equilibrium there are no genuine four-point correlations. 

Based on the general expressions given in \cite{MaySol04} we have verified 
that four-point correlations 
$\langle \s_k(t) \, \s_{l_1}(\tw) \, \s_{l_2}(\tw) \, \s_{l_3}(\tw) \rangle$ 
likewise factorize. There should be a generic connection between this 
property and the fact that the Glauber-Ising model maps to free
fermions~\cite{Felderhof71}. However, as we will see below, the 
factorization only holds in equilibrium. 

Now consider for comparison the standard four-point correlation function 
(\ref{equ:Cnregdef}). It can be expressed in terms of the connected 
one, Equation (\ref{equ:Cndef}), via 
\begin{eqnarray}
  \fl \C_{l-k}(\dt,\tw) = \Cc_{l-k}(\dt,\tw) \nonumber \\ 
    {}+{}\langle \s_k(t) \s_l(t) \rangle \langle \s_k(\tw) \s_l(\tw) \rangle 
    -\langle \s_k(t) \s_l(\tw) \rangle \langle \s_k(\tw) \s_l(t) \rangle. 
  \label{equ:Cconversion}
\end{eqnarray}
The first term vanishes in equilibrium, and so
$\C_n^{\eq}(\dt) = \C_n(\dt,\tw \to \infty)$ can be expressed purely in terms 
of {\em two-point} correlation functions.
This suggests that the standard four-point function $\C_{l-k}(\dt,\tw)$ 
is strongly biased by pairwise correlations and thus not suitable for 
revealing genuine four-point correlation effects.

\section{Non-Equilibrium}
\label{sec:1dneq}

The significance of measuring connected four-point correlations becomes 
even clearer when considering non-equilibrium coarsening dynamics. For the 
sake of simplicity we focus on a zero-temperature quench, i.e., $\gamma = 1$. 
The scaling of $\Cc_n(\dt,\tw)$ in the limit of large times $\dt,\tw \to 
\infty$ and distances $n \to \infty$ is then expected to be of the form 
\begin{equation}
  \Cc_n(\dt,\tw) \sim f_C\left(\frac{\dt}{2\tw},\frac{|n|}{2\sqrt{\tw}}\right).
  \label{equ:Cnscal}
\end{equation}
Formally, the long-time and long-distance limit is taken at fixed
values of the scaling variables 
\begin{equation}
  \alpha = \frac{\dt}{2\tw} \quad \mbox{and} \quad \eta =
\frac{|n|}{2\sqrt{\tw}}\ .
  \label{equ:an}
\end{equation}
The first of these, $\alpha$, measures the observation time interval $\dt$ 
in units of the system's age $\tw$ while $\eta$ is the ratio of distances 
$n$ over the typical domain size, bearing in mind that the latter
scales as $O\left( \sqrt{\tw} 
\right)$. The factors of $\frac{1}{2}$ in the definitions of
$\alpha,\eta$ are included for mathematical convenience in what follows.

\begin{figure}
  \hspace*{2.54cm} \epsfig{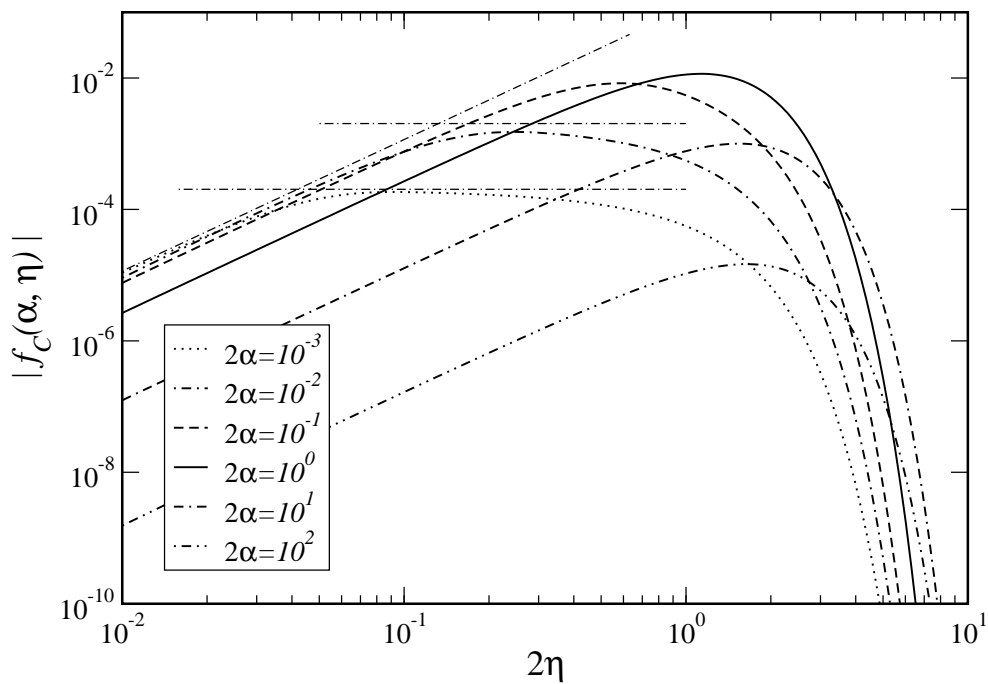}  
  \caption{\label{fig:CorVsEta} Modulus of the scaling function
    $f_C(\alpha,\eta)$ of the connected four-point correlation
    $\Cc_n(\dt,\tw)$ versus distance $2\eta=|n|/\sqrt{\tw}$ for various
    time ratios $2\alpha=\dt/\tw$. The curves are obtained by numerical
    evaluation of the exact scaling functions (\ref{equ:fCint}).
    The dashed-dotted straight lines are the asymptotes 
    (\ref{equ:fCrhoall1rll1}) and (\ref{equ:fCrhoall1rgg1}), 
    where the sloping line corresponds to (\ref{equ:fCrhoall1rll1}) 
    for the regime $\eta^2 \ll \alpha \ll 1$ while the horizontal ones 
    represent (\ref{equ:fCrhoall1rgg1}) with $2\alpha=10^{-3},10^{-2}$ 
    and apply in the regime $\alpha \ll \eta^2 \ll 1$. The data is 
    discussed below and interpreted in Section~\ref{sec:randomwalk}.}
\end{figure}

In order to obtain the scaling function $f_C(\alpha,\eta)$ a leading order 
asymptotic expansion of (\ref{equ:CnST}) is required. To this end we use the 
asymptotic formula~\cite{Mathbook}
\begin{equation}
  \rme^{-t} \, \In_q(t) \sim \frac{1}{\sqrt{2\pi t}} \, \rme^{-q^2/(2t)},
  \label{equ:Ilgn}
\end{equation}
which applies for $q,t \to \infty$ with $q^2/t$ fixed. In the same limit 
we have, by combining (\ref{equ:HI}) with (\ref{equ:Ilgn}) and
the identity $\In_{q-1}(t)-\In_{q+1}(t) = 
\frac{2q}{t} \In_q(t)$, and setting $\gamma=1$,
\begin{equation}
  \Hn_q(t) \sim \erfc\left( \frac{q}{\sqrt{2t}} \right)
  \quad \mbox{with} \quad 
  \erfc(x) = \frac{2}{\sqrt{\pi}} \int_x^\infty \rmd u \, \rme^{-u^2}. 
  \label{equ:Hasymp}
\end{equation}
The function $\erfc(x)$ is in fact just the complementary error function 
$\erfc(x) = \mathrm{erfc}(x) = 1-\mathrm{erf}(x)$; we use 
the symbol $\erfc$ to keep the notation compact. When substituting 
(\ref{equ:Ilgn}), (\ref{equ:Hasymp}) into (\ref{equ:Ssum}), (\ref{equ:Tsum}) 
the sums defining $\Scal_q, \Tcal_q$ turn into Riemann-sums such that 
\begin{eqnarray}
\Scal_q \sim \frac{2}{\sqrt{\pi}} \int_0^\infty \rmd x \, 
  \rme^{-x^2} \erfc\left( \frac{q}{2\sqrt{\tw}} + \sqrt{\frac{\dt}{2\tw}} 
  \, x \right),  
  \label{equ:Sint}\\
\Tcal_q \sim \frac{2}{\sqrt{\pi}} \int_0^\infty \rmd x \, 
  \rme^{-\left(x+n/\sqrt{2\dt}\right)^2} 
  \erfc\left( \frac{q}{2\sqrt{\tw}} + \sqrt{\frac{\dt}{2\tw}} \, x \right). 
  \label{equ:Tint}
\end{eqnarray}
Using the expansions (\ref{equ:Ilgn})-(\ref{equ:Tint}) in (\ref{equ:CnST}) 
and taking the scaling limit also turns the sums explicitly appearing in 
(\ref{equ:CnST}) into integrals. In terms of our scaling variables $\alpha, 
\eta$ the scaling function $f_C(\alpha,\eta)$ is thus
\begin{eqnarray}
  \fl f_C(\alpha,\eta) = 
    \frac{4}{\alpha\pi} \left\{ \int_0^\infty \! \rmd u \int_0^\infty \rmd v 
    \, \rme^{- \frac{1}{\alpha} \left[u^2 + \left(\eta + v \right)^2 \right]} 
    \left[ \erfc(\eta + u) \erfc(v) - \erfc(u) \erfc(\eta + v) \right] \right. 
    \nonumber \\
    + \;\!\int_0^\eta \, \rmd u \int_0^\infty \rmd v \, 
    \rme^{- \frac{1}{\alpha} \left[ u^2 + \left(\eta + v \right)^2 \right]} 
    \left[ \erfc(u+v) \erfc(\eta) - \erfc(u) \erfc(\eta + v) \right] 
    \nonumber \\
    - \left. \int_0^\eta \, \rmd u \int_0^\infty \rmd v \, 
    \rme^{-\frac{1}{\alpha} \left[ \left(\eta - u \right)^2 + v^2 \right]} 
    \left[ \erfc(u+v) \erfc(\eta) - \erfc(u) \erfc(\eta + v) \right] \right\}. 
  \label{equ:fCint}
\end{eqnarray}
Equation~(\ref{equ:fCint}) is suitable for numerical evaluation; the 
resulting plots of $f_C(\alpha,\eta)$ are shown inFig.~\ref{fig:CorVsEta}. 
One sees clearly that there are non-trivial four-point correlations in 
the non-equilibrium coarsening dynamics of the Glauber-Ising chain. We 
notice also that $f_C(\alpha,\eta)$ is {\em negative} throughout (the plot 
shows the modulus); we will find an explanation for this feature in 
Section~\ref{sec:randomwalk} below. Because of the rather rich structure 
in $f_C(\alpha,\eta)$ let us next discuss its various scaling regimes. 

First consider $\alpha \gg 1$, which corresponds to $\dt \gg \tw$. The
behaviour 
of $f_C(\alpha,\eta)$ in this regime is easily obtained form (\ref{equ:fCint}) 
by Taylor expanding the exponentials in $1/\alpha$. The zeroth order 
contributions, where the exponentials are replaced by unity, vanish: 
in the first integral in (\ref{equ:fCint}) the combination of $\erfc$'s 
is antisymmetric under 
exchanging $u,v$ while the second and third integral cancel. So the leading 
behaviour of $f_C(\alpha,\eta)$ follows from first-order contributions where, 
for the same reasons, various terms drop out. The remaining integrals, 
which are of the type $\int \rmd x \, x^i \, \erfc(x)$ with $i=0,1,2$ and 
$\int \rmd x \int \rmd y \, (x+y) \, \erfc(x+y)$, are solvable~\cite{Mathbook} 
and lead to the large $\alpha$ expansion 
\begin{eqnarray}
  \fl f_C(\alpha,\eta) = - \frac{2}{3\pi^2} \left\{ 12 \, \eta^2 \, 
  \rme^{-\eta^2} 
  -2 \left[ (1+6\eta^2) - (1 + \eta^2) \rme^{-\eta^2} \right] \sqrt{\pi} \, 
  \eta \, \erfc(\eta) \right. 
  \nonumber \\
  - (3+2\eta^2) \pi \, \eta^2 \, \erfc^2(\eta) \bigg\} \frac{1}{\alpha^2} 
  + O\left( \frac{1}{\alpha^3} \right).
  \label{equ:fCagg1}
\end{eqnarray}
This means that the magnitude of four-point correlations drops as
$1/\alpha^2$ for $\alpha \gg 1$, at all distances $\eta$. The plots of
the exact  
$f_C(\alpha,\eta)$ in Fig.~\ref{fig:CorVsEta} for $2\alpha=10^1,10^2$ 
illustrate 
this nicely. Details of the shape of $f_C(\alpha,\eta)$ at large $\alpha$ are 
revealed by expanding (\ref{equ:fCagg1}) for $\eta \ll 1$ and $\eta \gg 1$, 
i.e., 
distances $n$ far below and far above the typical domain-size at time $\tw$. 
In 
the former case we simply Taylor expand in $\eta$ while in the latter we 
use~\cite{Mathbook} that $\erfc(\eta) = 1/(\sqrt{\pi} \, \eta) \, 
\rme^{-\eta^2} [1 - 1/(2\eta^2) + O(\eta^{-4}) ]$. The leading terms are 
\begin{eqnarray}
  f_C(\alpha,\eta) \sim
    -\frac{2}{\pi} \left( \frac{4}{\pi} -1 \right) \frac{\eta^2}{\alpha^2} 
    & \quad \mbox{for} \quad \eta^2 \ll 1 \ll \alpha, 
    \label{equ:fCagg1nll1}
  \\
  f_C(\alpha,\eta) \sim 
    -\frac{8}{3\pi^2\alpha^2} \, \rme^{-\eta^2}
    & \quad \mbox{for} \quad 1 \ll \alpha, \eta^2. 
    \label{equ:fCagg1ngg1}
\end{eqnarray}
Again the plots in Fig.~\ref{fig:CorVsEta} for $2\alpha = 10^1,10^2$ clearly 
show the power-law behaviour given by (\ref{equ:fCagg1nll1}) and the Gaussian 
cutoff (\ref{equ:fCagg1ngg1}). 

Now we turn to $\alpha \ll 1$ or $\dt \ll \tw$. To obtain expansions of 
$f_C(\alpha,\eta)$ in this regime it is necessary to rearrange 
(\ref{equ:fCint}). It is further convenient to introduce 
\begin{equation}
  \hat{f}_C(\alpha,\rho) = f_C(\alpha,\sqrt{\alpha} \, \rho) 
  \quad \mbox{with} \quad \rho = \frac{\eta}{\sqrt{\alpha}} =
\frac{n}{\sqrt{2\dt}}\ . 
  \label{equ:fCrhoeta}
\end{equation}
The scaling variable $\rho$ may be viewed as an alternative measure for 
distance and thus replaces $\eta$. In this notation and by shifting and 
scaling the integration variables equation (\ref{equ:fCint}) can be 
re-expressed as  
\begin{eqnarray}
\fl \hat{f}_C(\alpha,\rho) = 
  \frac{4}{\pi} \left\{ \int_0^\infty \mathrm{D} u \int_\rho^\infty 
  \mathrm{D} v \, 
  \left[ \erfc_\alpha(\rho+u)\erfc_\alpha(v-\rho) 
  - \erfc_\alpha(u) \erfc_\alpha(v) \right] \right. 
  \nonumber \\ 
  + \int_0^\rho \mathrm{D} u \int_\rho^\infty \mathrm{D} v \, 
  \left[ \erfc_\alpha(v+u-\rho)\erfc_\alpha(\rho) 
  - \erfc_\alpha(u) \erfc_\alpha(v) \right] 
  \nonumber \\
  \left. \! - \int_0^\rho \mathrm{D} u \int_0^\infty \mathrm{D} v \, 
  \left[ \erfc_\alpha(v-u+\rho)\erfc_\alpha(\rho) 
  - \erfc_\alpha(\rho - u) \erfc_\alpha(\rho + v) \right] \right\}.
  \label{equ:fCrho}
\end{eqnarray}
Here we have introduced the short-hands $\mathrm{D}x = \rme^{-x^2} \rmd x$ 
and $\erfc_\alpha(x) = \erfc\left( \sqrt{\alpha} \, x \right)$. Equation 
(\ref{equ:fCrho}) is suitable for studying the small $\alpha$ regime of 
$f_C(\alpha,\eta)$. Two cases have to be distinguished: we can 
expand around $\alpha = 0$ either at fixed $\rho$ or at fixed $\eta$. 
When keeping $\rho$ fixed we effectively 
look at distances $\eta = \rho \, \sqrt{\alpha} = 
O\left( \sqrt{\alpha} \right)$, while obviously $\eta = O(1)$ if we fix 
$\eta$. Because the $\rho$ and $\eta$ length scales become disparate for 
$\alpha \to 0$, separate expansions must be made. 

The shape of $f_C(\alpha,\eta)$ for small $\alpha$ and fixed $\rho$ 
immediately follows from (\ref{equ:fCrho}) by Taylor expanding the functions 
$\erfc_\alpha(x) = 
\erfc\left(\sqrt{\alpha} \, x\right)$ in $\sqrt{\alpha}$. This turns the 
integrands in (\ref{equ:fCrho}) into Gaussians (contained in $\mathrm{D}u, 
\mathrm{D}v$) with polynomial factors. Evaluating the integrals gives 
\begin{equation}
\fl\hat{f}_C(\alpha,\rho) = - \frac{4}{\pi^2} \left\{ 1 - 
  \left[ \rme^{-\rho^2} + \sqrt{\pi} \, \rho \left( 2 - \erfc(\rho) \right) 
  \right] 
  \left[ \rme^{-\rho^2} - \sqrt{\pi} \, \rho \, \erfc(\rho) \right] 
  \right\} \alpha + O\left( \alpha^{3/2} \right). 
  \label{equ:fCrhoall1}
\end{equation}
To understand the result (\ref{equ:fCrhoall1}) it is instructive to consider 
the limits $\rho \ll 1$ and $\rho \gg 1$, corresponding to distances $n \ll 
\sqrt{\dt}$ and $n \gg \sqrt{\dt}$, respectively. One finds to leading order 
and in terms of $\alpha$ and $\eta$ 
\begin{eqnarray}
  f_C(\alpha,\eta) \sim - \frac{4}{\pi} \left( 1 - \frac{2}{\pi} \right) 
  \eta^2 
  & \quad \mbox{for} \quad \eta^2 \ll \alpha \ll 1,
  \label{equ:fCrhoall1rll1} \\
  f_C(\alpha,\eta) \sim - \frac{4}{\pi^2} \, \alpha 
  & \quad \mbox{for} \quad \alpha \ll \eta^2 \ll 1. 
  \label{equ:fCrhoall1rgg1} 
\end{eqnarray}
We note that at small $\rho$ the leading term in (\ref{equ:fCrhoall1}) is 
$\rho^2 \alpha = \eta^2$ and thus (\ref{equ:fCrhoall1rll1}) follows. So 
$f_C(\alpha,\eta)$ initially grows as $\eta^2$ independently of $\alpha$. 
But for $\eta \approx \sqrt{\alpha}$ four-point correlations 
level off at a plateau of height $O(\alpha)$, as given by equation 
(\ref{equ:fCrhoall1rgg1}).
The asymptotes (\ref{equ:fCrhoall1rll1}), (\ref{equ:fCrhoall1rgg1}) are shown 
in Fig.~\ref{fig:CorVsEta}.

It remains to discuss the behaviour of $f_C(\alpha,\eta)$ for small
$\alpha$ and fixed $\eta$.
In this case $\rho = \eta/\sqrt{\alpha}$ in (\ref{equ:fCrho}) diverges for 
$\alpha \to 0$. Because of the Gaussian 
weights in $\mathrm{D}u, \mathrm{D}v$ only integrals containing the 
neighborhood 
of $u=v=0$ then contribute significantly to $f_C(\alpha,\eta)$. This
holds 
for the third integral in (\ref{equ:fCrho}), but not for the other two. 
The first one, for instance, satisfies the bound
\begin{equation*}
  \left| \frac{4}{\pi} \int_0^\infty \mathrm{D} u \int_\rho^\infty 
  \mathrm{D} v \, 
  \left[ \erfc_\alpha(\rho+u)\erfc_\alpha(v-\rho) 
  - \erfc_\alpha(u) \erfc_\alpha(v) \right] \right| \leq 
  2 \erfc(\eta)\erfc(\rho).
\end{equation*}
This follows from the triangular inequality, the identities $\erfc(\rho) = 
(2/\sqrt{\pi}) \int_\rho^\infty \mathrm{D}x$ and $\erfc(\eta)=
\erfc_\alpha(\rho)$ 
and the fact that $\erfc(x)$ is monotonously decreasing. The same bound 
can be used for the second integral in (\ref{equ:fCrho}).
Extending 
the $u$-integration range in the third integral to $\int_0^\infty \mathrm{D}u$ 
likewise only produces excess contributions of the same size. Therefore, to 
order $O\left(\erfc(\eta)\erfc(\rho)\right)$ equation (\ref{equ:fCrho}) 
reduces to 
$$
  f_C(\alpha,\eta) \simeq - \frac{4}{\pi} 
  \int_0^\infty \mathrm{D} u \int_0^\infty \mathrm{D} v \, 
  \left[ \erfc\left(\sqrt{\alpha}(v-u)+\eta\right)\erfc\left(\eta\right) 
  - \erfc\left(\eta - \sqrt{\alpha}u\right) \erfc\left(\eta + 
  \sqrt{\alpha}v\right) \right] 
$$
at small $\alpha$ and fixed $\eta$. Here we have substituted $\erfc_\alpha(x) 
= \erfc\left(\sqrt{\alpha} \, x\right)$ and $\sqrt{\alpha} \, \rho = \eta$. As 
will become clear in a moment the above expression has power-law
scaling at small $\alpha$. On the other hand, $\erfc(\rho) \sim
\sqrt{\alpha}/(\sqrt{\pi} \, \eta) \,
\rme^{-\eta^2/\alpha}$ vanishes faster than any power-law for $\alpha \to 0$. 
Therefore we may safely ignore the $O\left(\erfc(\eta)\erfc(\rho)\right)$ 
contributions discarded above. By Taylor expanding the last representation 
for $f_C(\alpha,\eta)$ in $\sqrt{\alpha}$, which leads to simple Gaussian 
integrals, we thus finally obtain the small $\alpha$ scaling at fixed $\eta$, 
\begin{equation}
  f_C(\alpha,\eta) = - \frac{4}{\pi^2} \rme^{-\eta^2} \left[ \rme^{-\eta^2} 
  -  \sqrt{\pi} \, \eta \, \erfc(\eta) \right] \alpha + O\left( \alpha^{3/2} 
  \right). 
  \label{equ:fCetaall1}
\end{equation}
Equation (\ref{equ:fCetaall1}) tells us that, on the length
scale set by $\eta$, four-point  
correlations decrease linearly with $\alpha$ as $\alpha\to 0$, for any
value of $\eta$. The plots for $2\alpha=10^{-2}, 
10^{-3}$ in Fig.~\ref{fig:CorVsEta} illustrate this. At small and
large $\eta=|n|/(2\sqrt{\tw})$
the behaviour of (\ref{equ:fCetaall1}) is to leading order 
\begin{eqnarray}
  f_C(\alpha,\eta) \sim - \frac{4}{\pi^2} \alpha 
  & \quad \mbox{for} \quad \alpha \ll \eta^2 \ll 1,
  \label{equ:fCetaall1ell1} \\
  f_C(\alpha,\eta) \sim - \frac{2\alpha}{\pi^2\eta^2} \, \rme^{-2\eta^2}
  & \quad \mbox{for} \quad \alpha \ll 1 \ll \eta^2. 
  \label{equ:fCetaall1egg1}
\end{eqnarray}
The plateau (\ref{equ:fCetaall1ell1}) obtained at small $\eta$ 
matches the one at large $\rho$, Equation (\ref{equ:fCrhoall1rgg1}),
as it should. 
For small $\alpha$ and large $\eta$, finally, $f_C(\alpha,\eta)$ has an 
essentially Gaussian cutoff (\ref{equ:fCetaall1egg1}). The latter occurs 
at slightly smaller $\eta$ than in the large $\alpha$ case, see 
(\ref{equ:fCagg1ngg1}) and Fig.~\ref{fig:CorVsEta}.

\begin{figure}
  \hspace*{2.54cm} \epsfig{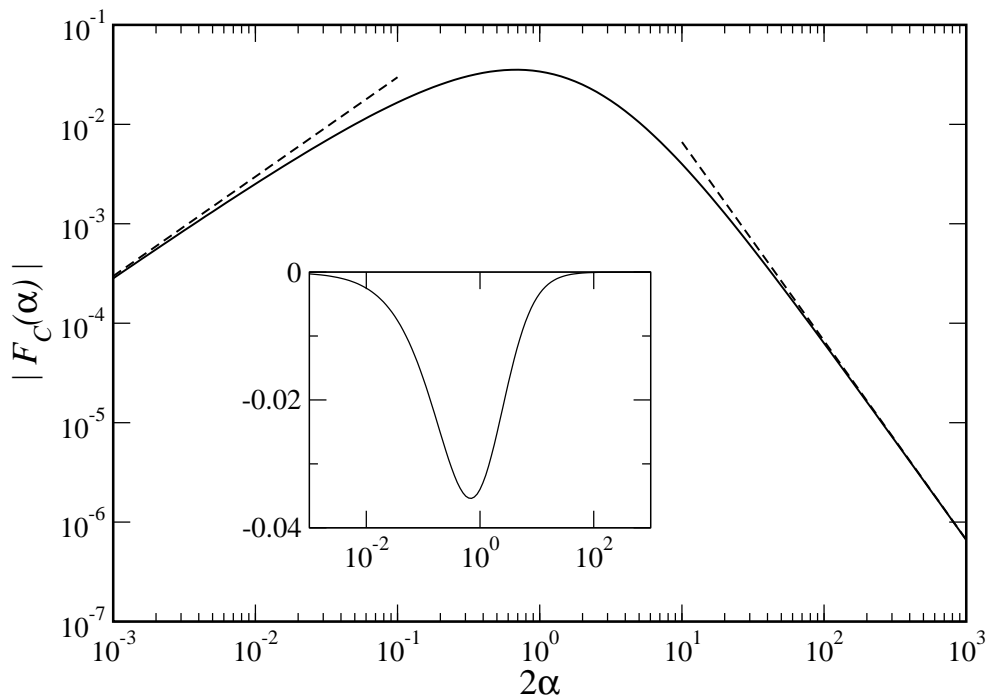} 
  \caption{\label{fig:CorSum} Modulus of the scaling function
    $F_C(\alpha)$ of the four-point susceptibility $\Xc(\dt,\tw)$
    defined in (\ref{equ:FC}), versus $2\alpha=\dt/\tw$ for
    zero temperature coarsening.  Dashed lines represent the asymptotes 
    (\ref{equ:Fageall1}) and (\ref{equ:Fageagg1}). Inset: $F_C$
    on a linear scale.}
\end{figure}

Our discussion of $f_C(\alpha,\eta)$ has given us a complete understanding 
of the spatio-temporal scaling of the connected four-point correlation 
$\Cc_n(\dt,\tw)$. The various scaling regimes in Fig.~\ref{fig:CorVsEta} 
are characterized by (\ref{equ:fCagg1nll1}), (\ref{equ:fCagg1ngg1}), 
(\ref{equ:fCrhoall1rll1}), (\ref{equ:fCrhoall1rgg1}), 
(\ref{equ:fCetaall1ell1}) 
and (\ref{equ:fCetaall1egg1}). We now conclude our analysis of the connected 
four-point correlation in nonequilibrium coarsening by studying the scaling 
of the associated four-point susceptibility. From its definition 
(\ref{equ:Cdef}), 
the properties $\Cc_{-n}(\dt,\tw) = \Cc_n(\dt,\tw)$, $\Cc_0(\dt,\tw)=0$ and 
the scaling (\ref{equ:Cnscal}) one finds in the large-time limit
$\dt,\tw \to \infty$ at fixed $\alpha$, 
\begin{equation*}
  \fl
  \Xc(\dt,\tw) = 2 \sum_{n=1}^\infty \Cc_n(\dt,\tw) 
  \sim 2 \sum_{n=1}^\infty f_C\left(\frac{\dt}{2\tw},\frac{n}{2\sqrt{\tw}}
  \right) 
  \sim 4\sqrt{\tw} \int_0^\infty \rmd \eta \, f_C(\alpha,\eta).
\end{equation*}
This defines the scaling function $F_C(\alpha)$ for the susceptibility, 
\begin{equation}
  \Xc(\dt,\tw) \sim \sqrt{\tw} \, F_C\left(\frac{\dt}{2\tw}\right) 
  \quad \mbox{with} \quad 
  F_C(\alpha) = 4 \int_0^\infty \rmd \eta \, f_C(\alpha,\eta).
  \label{equ:FC}
\end{equation}
Numerical integration of (\ref{equ:FC}) using $f_C(\alpha,\eta)$  as given 
in (\ref{equ:fCint}) produces the plots in Fig.~\ref{fig:CorSum}. Because 
$f_C(\alpha,\eta)<0$ we also find that $F_C(\alpha)$ is negative throughout. 
Furthermore, $F_C(\alpha)$ shows power-law behavior at small as well as large 
$\alpha$: 
\begin{eqnarray}
  F_C(\alpha) = -\frac{8\left(\sqrt{2}-1\right)}{\pi^{3/2}}
    \, \alpha + O(\alpha^{3/2}) 
    & \quad \mbox{for} \quad \alpha \ll 1,  
  \label{equ:Fageall1} \\
  F_C(\alpha) = -\frac{2\left(8\sqrt{2}-9\right)}{5\pi^{3/2}} 
    \, \frac{1}{\alpha^2} + O\left(\frac{1}{\alpha^3}\right) 
    & \quad \mbox{for} \quad \alpha \gg 1. 
  \label{equ:Fageagg1}
\end{eqnarray}
The expansion for large $\alpha$ is obtained by substituting the corresponding 
expansion (\ref{equ:fCagg1}) of $f_C(\alpha,\eta)$ in (\ref{equ:FC}). At small 
$\alpha$ we split $\int_0^\infty \rmd \eta = \int_0^{\nu \sqrt{\alpha}} 
\rmd \eta 
+ \int_{\nu \sqrt{\alpha}}^\infty \rmd \eta$ in (\ref{equ:FC}) with some 
$\nu > 0$. 
The integrals  correspond to the $\rho$ and $\eta$ length scales where the 
expansions 
(\ref{equ:fCrhoall1}) and (\ref{equ:fCetaall1}) of $f_C(\alpha,\eta)$ apply, 
respectively. One easily shows that the contributions to $F_C(\alpha)$
from the  
$\rho$ length scale are $O(\alpha^{3/2})$ while those from the $\eta$
scale grow as $O(\alpha)$. Therefore the leading term in the small $\alpha$ 
expansion 
(\ref{equ:Fageall1}) is given by substituting (\ref{equ:fCetaall1}) into
(\ref{equ:FC}). The fact that $F_C(\alpha)$ vanishes at small and large 
$\alpha$ 
is not surprising: as discussed below (\ref{equ:Cndef}) the connected 
four-point 
correlation $\Cc_n(\dt,\tw)$ goes to zero for $\dt \to 0$ as well as 
$\dt \to \infty$. 
However, that the approach to these limits is through the power laws
(\ref{equ:Fageall1}) and (\ref{equ:Fageagg1}) is rather non-trivial.
From the arguments above it is clear that the linear scaling at small
$\alpha$ is a consequence of the same scaling of the plateau height in
$f_C(\alpha,\eta)$. We consider next a physical picture which provides
some intuition into how this scaling arises.

\section{Random Walk Interpretation}
\label{sec:randomwalk}

The exact scaling results summarized in Figure~\ref{fig:CorVsEta} have a 
lot of structure. To develop a physical understanding, we now use the 
fact that low-temperature dynamics of the Ising chain can be described
in terms of domain walls that perform independent random 
walks with diffusion rate $\frac{1}{2}$ until they meet, 
when they annihilate with rate close to unity~\cite{Santos97}. In
equilibrium there is also  
the reverse process, where pairs of domain walls or ``walkers'' are created 
at a small rate.

\begin{figure}
  \hspace*{2.54cm} \epsfig{file=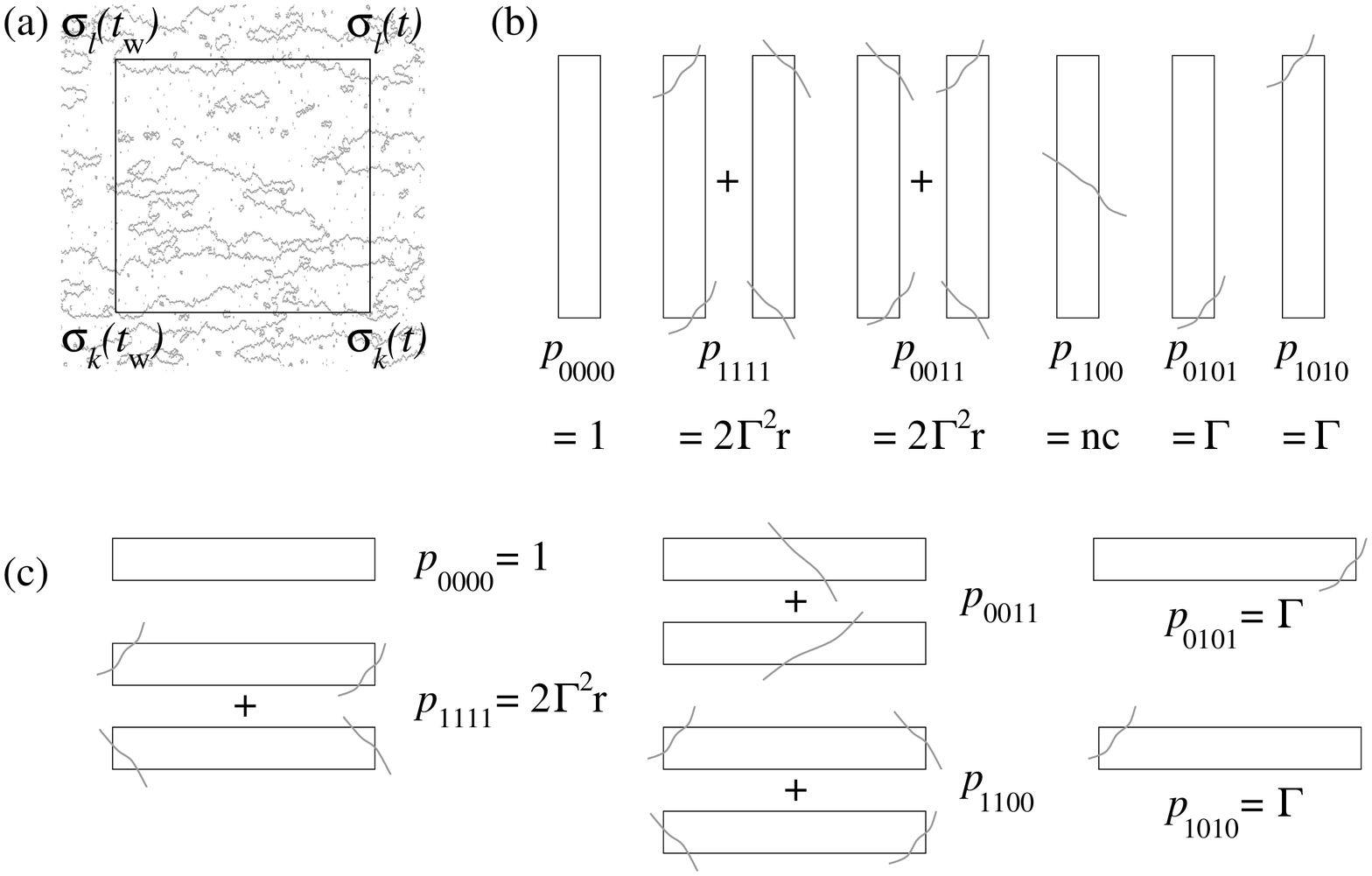,width=13cm}
\caption{Space-time representation of the connected four-point
correlation; grey lines indicate domain-wall trajectories~\cite{GC02}. 
Panel (a): snapshot of trajectories over a spatial region (vertical) 
of 500 sites and a time-window (horizontal) of $\dt=1500$ for equilibrium 
dynamics at $T=0.6$. Panel (b), (c): schematic trajectories. 
See text for discussion.
\label{fig:spacetime}}
\end{figure}

In a space-time diagram~\cite{GC02}, Figure~\ref{fig:spacetime}(a), 
the spins
$\sigma_k(\tw)$, $\sigma_l(\tw)$, $\sigma_k(t)$ and $\sigma_l(t)$ 
define the four corners of a rectangle. Spin products are then determined
by whether an even or odd number of domain walls cross the relevant
edge of the rectangle. For instance $\sigma_k(\tw)\sigma_k(t)$ equals $1$ 
if an even number of walkers cross the bottom edge; otherwise it equals
$-1$. One can then classify all possible situations by the parity of
the number of walkers crossing the four edges. Numbering the edges in
the order left - right - top - bottom, we will for example denote the
situation where an odd number of walkers crosses on the left and
bottom as 1001. Because walkers only annihilate or recreate in pairs,
the total number of walkers crossing the rectangle has to be even, so
that there are eight possible situations. After a short calculation
one shows that, in terms of the corresponding probabilities, the
connected four-point correlation is
\begin{equation}
\Cc_n(\dt,\tw) = 8 (p_{0000}\,p_{1111}+p_{0011}\,p_{1100} - p_{0101}\,p_{1010}
- p_{0110}\,p_{1001}).
\label{equ:Cn_in_ps}
\end{equation}
The last two terms are always identical to each other, due to the 
spatial mirror symmetry of the problem. The representation
(\ref{equ:Cn_in_ps}) is particularly useful when the number of walkers
crossing the rectangle is small. We expect this to be the case in the
regime $n\ll 1/c(\tw)$ and $\sqrt{\dt}\ll 1/c(\tw)$, where $c(\tw)$ is
the concentration of domain walls at time $\tw$. In the coarsening regime,
where $c(\tw)\sim 1/(2\sqrt{\pi \, \tw})$, these
conditions translate to $n^2\ll \tw$ and $\dt\ll\tw$. In equilibrium,
on the other hand, $c$ is a constant fixed by the temperature. To
include both cases, we simply write $c$ below, even for the coarsening case.

Our restrictions so far still leave open the relative magnitude of
$\dt$ and $n^2$; let us focus first on the case where $\dt\ll
n^2$. The space-time rectangle is then wide in the space-direction and
narrow in the time-direction. From this one can deduce the leading
contributions to the various probabilities, which are shown in
Figure~\ref{fig:spacetime}(b).  We have denoted by $\Gamma$ the
probability that a walker will cross a corner of the rectangle. Since
$n\gg \sqrt{\dt}$, this probability is dominated by the smallness of
$\sqrt{\dt}$, while the spatial extent of the box is
irrelevant. Thus, $\Gamma$ can be calculated as the probability that a
walker will cross from one half-space into the other during the time
interval $\dt$, which is easily found as $\Gamma =
c\sqrt{\dt/2\pi}$. The other quantity that appears in the
probabilities is the factor $r$, which we define to be the joint
probability at time $\tw$ of having two walkers at distance $n$,
normalized by $c^2$. With this normalization, one would have $r=1$ for
$n\gg 1/c$ because correlations between walkers vanish at large
distance. At equilibrium, where walkers are uncorrelated at {\em any}
distance, one in fact has $r=1$ even for $n\ll 1/c$. In the coarsening 
situation, on the other hand, $r$ is of order $nc$ for $n\ll 1/c$ and
thus vanishes to leading order. This reflects the effective repulsion
between walkers: a walker that has survived the coarsening dynamics up to
$\tw$ is not likely to have other walkers within a distance of order
$O(\sqrt{\tw}) = O(1/c)$. Putting the terms from
Figure~\ref{fig:spacetime}(b) together gives
\begin{equation}
\fl \Cc_n(\dt,\tw) \approx 8 (1\times 2\Gamma^2 r+ 2\Gamma^2 r\times nc
-2 \Gamma^2) \approx 16 \Gamma^2 (r-1) = \frac{8}{\pi}(r-1) c^2 \dt, 
\label{equ:Cn2}
\end{equation}
where we have used the fact that $nc\ll 1$ to neglect the second
term. This simple expression explains two important qualitative
observations made above. First, in equilibrium, $\Cc_n$ {\em vanishes} because
$r=1$. Second, in the coarsening case, $\Cc_n$ is {\em negative} because
$r<1$ when $n \ll 1/c$. Thus, the sign of $\Cc_n$ arises from the effective
repulsion between walkers discussed previously. Quantitatively, using that
$r\ll 1$ for $n\ll 1/c$, the result (\ref{equ:Cn2}) predicts that 
\begin{equation}
  \Cc_n \approx - \frac{8}{\pi} \, c^2 \dt \sim - \frac{2}{\pi^2} \, 
  \frac{\dt}{\tw} = - \frac{4}{\pi^2} \, \alpha. 
  \label{equ:Cn2.1}
\end{equation} 
This is precisely our expansion (\ref{equ:fCrhoall1rgg1}) for the 
regime $\dt\ll n^2\ll \tw$, where $\Cc_n$ has an $n$-independent 
plateau whose height increases linearly with $\dt/\tw$. So we now 
have a microscopic picture for the occurrence of this plateau in 
terms of domain-wall dynamics. 

Next we apply similar arguments to the regime where $n$ is small
compared to $\sqrt{\dt}$, $n^2 \ll \dt \ll \tw$. As shown in
Figure~\ref{fig:spacetime}(c), the space-time rectangle is now extended
in the time-direction. As a consequence, the two probabilities
$p_{0011}$ and $p_{1100}$ swap their leading contributions.  For
$p_{0011}$, the leading term is now produced by a single walker
crossing the rectangle from top to bottom or from bottom to top. Since
$n\ll \sqrt{\dt}$, the spatial width of the box can be neglected to
leading order, and $p_{0011}$ reduces to the probability of crossing
from one halfspace to the other. Bearing in mind that the crossing can
occur from the top or the bottom then gives $p_{0011} =
2c\sqrt{\dt/2\pi}$. For $p_{1100}$, one might naively expect 
to get a product of two corner-crossing probabilities times a repulsion
factor. However, as opposed to the case of $p_{1111}$, the two corner
crossings shown can in fact be achieved by a {\em single} walker which
starts and ends within the interval of size $n$. This gives a leading
contribution of $p_{1100} = nc\times n/\sqrt{2\pi \dt}$. The
remaining terms are calculated as before, except for the fact that now
the corner-crossing probability is $\Gamma = nc/2$.  Assembling all
terms, and using again that $r\ll 1$ to neglect $p_{0000} \, p_{1111} = 
2 \Gamma^2 r$, we thus get for the coarsening case
\begin{equation}
\Cc_n(\dt,\tw) \approx 8 \left[2c\sqrt{\frac{\dt}{2\pi}} \times 
\frac{n^2c}{\sqrt{2\pi \dt}}-2 \Gamma^2\right] = 
-4 \left(1-\frac{2}{\pi}\right) n^2 c^2.
\label{equ:Cn3}
\end{equation}
This again has the correct negative sign overall. It also predicts
that, in this small-$n$ regime, $\Cc_n$ grows quadratically with $n$,
with an amplitude independent of $\dt$. In fact, using $n^2 c^2 \sim 
n^2/(4 \pi \tw) = \eta^2/\pi$, the result (\ref{equ:Cn3}) coincides 
with the expansion (\ref{equ:fCrhoall1rll1}) as it should.

The random walk picture has turned out to be useful for explaining the 
behaviour of $\Cc_n$ when $n^2, \dt \ll \tw$. In the remaining 
regimes discussed in the previous Section, on the other hand, 
where either $n^2$ or 
$\dt$ are large compared to $\tw$, it is less helpful because
a large number of annihilating walkers 
has to be considered. It is then no longer obvious how to estimate 
the probabilities in (\ref{equ:Cn_in_ps}). Nevertheless, the Gaussian 
cutoff for $n^2 \gg \tw$ that we found in (\ref{equ:fCagg1ngg1}), 
(\ref{equ:fCetaall1egg1}) is at least qualitatively reasonable: 
for length scales $n \gg \sqrt{\tw}$, 
correlations between random walkers become weak and one should 
effectively retrieve the equilibrium situation, where $\Cc_n=0$.

\section{Standard Functions out of Equilibrium}
\label{sec:regnoneq}

\begin{figure}
  \hspace*{2.54cm} 
  \epsfig{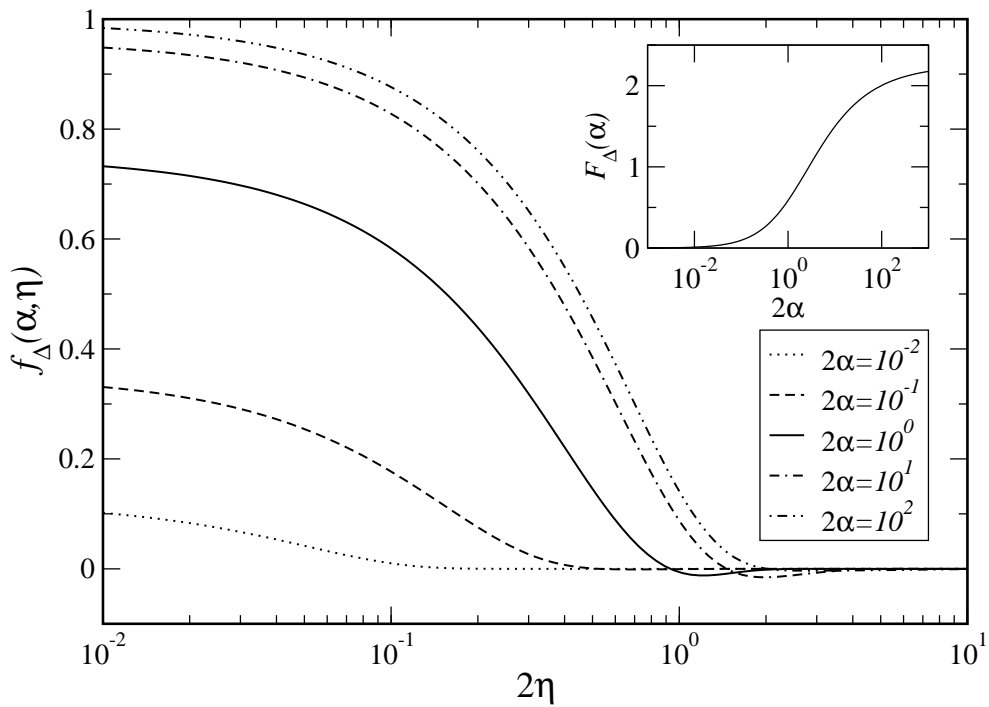}
  \caption{\label{fig:deltaC} Plots of the scaling expansion 
    (\ref{equ:dCn}) of $\Delta C_n(\dt,\tw)$ versus scaled distance 
    $2\eta=|n|/\sqrt{\tw}$ for various time ratios $2\alpha=\dt/\tw$. 
    Inset: Normalized contribution $F_\Delta(\alpha)$ 
    of the two-spin terms to the four-point susceptibility 
    $\X(\dt,\tw)$, Equation~(\ref{equ:dC}).}
\end{figure}

In Section~\ref{sec:1deq} we saw that while the connected four-point 
correlation $\Cc_n(\dt,\tw)$ and its associated 
four-point susceptibility $\Xc(\dt,\tw)$ vanish in equilibrium, the 
standard functions $\C_n(\dt,\tw)$ and $\X(\dt,\tw)$ are biased 
by two-spin correlations. It is the purpose of this section to show 
that the same is true for the non-equilibrium coarsening dynamics.
The link (\ref{equ:Cconversion}) between the standard and 
connected four-point correlations allows us to express their difference 
$\Delta C_n(\dt,\tw) = \C_n(\dt,\tw) - \Cc_n(\dt,\tw)$ purely in 
terms of two-spin correlations. This makes the analysis of 
$\Delta C_n(\dt,\tw)$ rather simple: spatial correlations are given 
in (\ref{equ:structfact}) in terms of $\Hn_n$ while temporal 
correlations for zero temperature coarsening have the exact 
representation \cite{MaySol04} 
\begin{equation}
  \fl \langle \s_k(t)\s_l(\tw) \rangle = \rme^{-(t+\tw)} \left\{ 
    \In_n(t+\tw)+\int_0^{2\tw} \rmd\tau \, 
    \In_n(t+\tw-\tau)
    \left[\In_0(\tau)+\In_1(\tau) \right] \right\}. 
  \label{equ:cor11zerotemp}
\end{equation}
In the scaling limit $\dt,\tw,n \to \infty$ with $\alpha=\dt/(2\tw)$ and 
$\eta=|n|/(2\sqrt{\tw})$ fixed we substitute the expansion (\ref{equ:Ilgn}) 
into (\ref{equ:cor11zerotemp}).
Combining terms according to (\ref{equ:Cconversion}) and some rearranging 
then produces $\Delta C_n(\dt,\tw) \sim f_\Delta(\alpha,\eta)$ with 

\begin{equation}
  \fl f_\Delta(\alpha,\eta) =
    \erfc(\eta) \, \erfc\left(\frac{\eta}{\sqrt{1+2\alpha}}\right) - 
    \left\{ \frac{2}{\pi} \int_0^{\arccot\sqrt{\alpha}} \rmd z \, 
    \rme^{-[\eta^2/(1+\alpha)] \, \sec^2(z)} \right\}^2,  
  \label{equ:dCn}
\end{equation}
where $\sec(z) = 1/\cos(z)$. The scaling of the difference 
between the four-point susceptibilities $\Delta \chi(\dt,\tw) = \X(\dt,\tw) 
- \Xc(\dt,\tw) = \sum_n \Delta C_n(\dt,\tw)$ then follows by analogy
with 
(\ref{equ:FC}): writing $\Delta \chi(\dt,\tw) \sim \sqrt{\tw} \, 
F_\Delta(\alpha)$ 
we obtain via integration of (\ref{equ:dCn}) over $\eta$, 
\begin{equation}
  \fl F_\Delta(\alpha) \sim \frac{4}{\sqrt{\pi}} \left\{ 
    \sqrt{1+2\alpha}-1
  + \frac{4}{\pi} \left[ 
    \arctan\sqrt{2\alpha} - \sqrt{\frac{1+\alpha}{2}} \, 
    \arctan2\sqrt{\alpha (1+\alpha)} \right] \right\}. 
  \label{equ:dC}
\end{equation}
Plots of $f_\Delta(\alpha,\eta)$ and $F_\Delta(\alpha)$ are 
shown in Figure~\ref{fig:deltaC}. Comparing the vertical scales in
Figures~\ref{fig:CorVsEta},~\ref{fig:CorSum},~\ref{fig:deltaC} demonstrates 
that the standard functions $\C_n(\dt,\tw)$ and $\X(\dt,\tw)$ are
completely dominated by the two-spin contributions (\ref{equ:dCn}),
(\ref{equ:dC}). A plot of the four-point susceptibility $\X_n(\dt,\tw) =
\Xc_n(\dt,\tw) + \Delta \X_n(\dt,\tw) \sim \sqrt{\tw} \left[ F_C(\alpha) 
+ F_\Delta(\alpha) \right]$, for instance, would be
indistinguishable by eye from the inset of Figure~\ref{fig:deltaC}. 

Therefore, as
claimed, the standard four-point function (\ref{equ:Cnregdef}) and its
associated four-point susceptibility (\ref{equ:Cdef}) are not suitable
for measuring genuine four-point correlations in the coarsening
dynamics of the Glauber-Ising chain. In comparison to strongly 
heterogeneous systems the relative magnitudes of, e.g., the connected 
four-point susceptibility (inset in Figure~\ref{fig:CorSum}) and the 
corresponding two-point bias (inset in Figure~\ref{fig:deltaC}) are 
reversed in coarsening systems.

\section{Conclusions}
\label{sec:conc}

In this paper we have explored dynamical heterogeneities in coarsening
systems by studying multi-point correlations in the dynamics of the
Glauber-Ising chain.  Since conventional four-point correlation
functions become dominated by strong spatial correlations that develop
in coarsening systems at late times, we considered ``connected''
four-point functions where these uninteresting two-point contributions
are eliminated.  We were able to obtain exact results and scaling
forms for these functions and the associated spatial integral, i.e.\ the
connected four-point susceptibility.  As a function of the time
difference $\dt$, this multi-point susceptibility has an extremum, as
is found in glass formers, for times of the order of the waiting time,
indicating the timescale for which dynamic heterogeneity is maximal.

Interestingly, we found that the connected four-point susceptibility
is negative throughout, and we were able to give an interpretation for
this behaviour in terms of the dynamics of domain walls, which undergo
free diffusion and pair annihilation.  The negative sign of the
susceptibility directly reflects the fact that there is an effective
repulsion between the domain walls, each having a ``depleted zone''
around it where the likelihood of finding another domain wall is low.
At equilibrium, on the other hand, domain wall positions are
uncorrelated and this leads to the vanishing of the susceptibility,
and of the underlying four-point correlations, for all $\dt$.  This
latter result, which we established using explicit expressions for
four-spin correlations, appears rather non-trivial.  It would be
interesting to verify whether it also extends to equilibrium
correlation functions of higher order.  If it does, one suspects that
there should be a deeper reason, possibly related to the mapping of
the Glauber-Ising chain dynamics to free fermions~\cite{Felderhof71}.

We also discussed the spatial dependence of the connected four-point
correlation functions. This has a richer structure than one might have
expected, but the random walk picture again gave a good qualitative
(and, in some regimes, quantitative) understanding of our exact
results.  In future work, it will be interesting to see if and how our
findings generalize to other coarsening systems. Encouragingly,
simulations show that many of the key features we found here extend at
least to two-dimensional Ising models~\cite{PeterLetter}.

\ack
We acknowledge financial support from the Austrian Academy of Sciences 
and EPSRC Grant No.\ 00800822 (PM), and EPSRC Grants No.\ GR/R83712/01, 
GR/S54074/01 and the University of Nottingham Grant No.\ FEF 3024 (JPG).

\section*{References}

\bibliography{DHQ}

\end{document}